\documentclass[]{lipics-v2021}
\usepackage[utf8]{inputenc}
\usepackage{color}

\hideLIPIcs \nolinenumbers

\usepackage{hyperref}
\usepackage{graphicx}

\title{Shadoks Approach to Knapsack Polygonal Packing}

\supplementdetails[subcategory={Source Code}]{}{https://github.com/gfonsecabr/shadoks-CGSHOP2024}

 		\author{Guilherme D. da Fonseca}
 	{LIS, Aix-Marseille Université}
 	{guilherme.fonseca@lis-lab.fr}
 	{https://orcid.org/0000-0002-9807-028X}
 	{}
 		\author{Yan Gerard}
 	{LIMOS, Université Clermont Auvergne}
 	{yan.gerard@uca.fr}
 	{https://orcid.org/0000-0002-2664-0650}
 	{}
 	
 	\authorrunning{Guilherme D. da Fonseca and Yan Gerard}

\funding{Work supported by the French ANR PRC grant ADDS (ANR-19-CE48-0005).}

\Copyright{Guilherme D. da Fonseca and Yan Gerard}

\acknowledgements{We would like to thank the CG:SHOP Challenge organizers and other competitors for their time, feedback, and making this whole event possible. We would like to thank Hélène Toussaint, Raphaël Amato, Boris Lonjon, and William Guyot-Lénat from LIMOS, as well as the Qarma and TALEP teams and Manuel Bertrand from LIS, who continue to make the computational resources of the LIMOS and LIS clusters available to our research. We would also like to thank Aldo Gonzalez-Lorenzo and the undergraduate students Aymeric Beck, Houssam Boufarachan, Marine Izoulet, and Carla Scardigli for coding viewers for the solutions.}

\begin{document}

\maketitle

\begin{abstract}
The 2024 edition of the CG:SHOP Challenge focused on the knapsack polygonal packing problem. Each instance consists of a convex polygon known as the \textit{container} and a multiset of \textit{items}, where each item is a simple polygon with an associated integer \textit{value}. A feasible packing solution places a selection of the items inside the container without overlapping and using only translations. The goal is to achieve a packing that maximizes the total value of the items in the solution.

Our approach to win first place is divided into two main steps. First, we generate promising initial solutions using two strategies: one based on integer linear programming and the other one employing a combination of geometric greedy heuristics. In the second step, we enhance these solutions through local search techniques, which involve repositioning items and exploring potential replacements to improve the total value of the packing.
\end{abstract}

\begin{figure}[b]
 \centering
 \includegraphics[height=3.5cm]{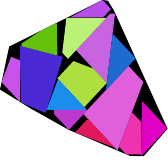} \hspace{.1em}
 \includegraphics[height=3.5cm]{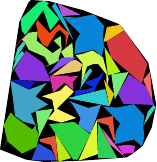} \hspace{.1em}
 \includegraphics[height=4cm]{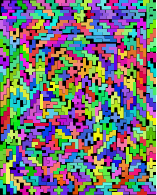} \hspace{.5em}
 \includegraphics[height=4cm]{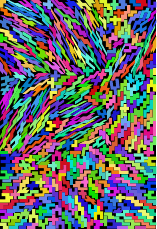}
 \caption{Our best solutions to \texttt{jigsaw\_cf2\_5db5d75a\_34}, \texttt{random\_rcf4\_6e323d40\_100},  \texttt{atris1240}, and \texttt{satris1786} instances.}
 \label{f:sols}
\end{figure}

\section{Introduction} \label{s:intro}

The CG:SHOP Challenge is an annual competition in geometric optimization. In its sixth edition in 2024, the challenge focuses on a two-dimensional knapsack packing problem. Our team, called \textit{Shadoks}, won first place with the best solution (among the 14 participating teams) to $75$ instances out of $180$ instances.

In this paper, we outline the heuristics we employed, beginning with a brief overview of the problem. Each input instance comprises a convex polygon referred to as the \textit{container} and a multiset of \textit{items}, where each item is a simple polygon associated with an integer \textit{value}. The objective is to pack a selection of the items inside the container using integer translations, maximizing the total sum of their values.

A total of 180 instances were provided, ranging from $28$ to $50{,}000$  items. The instances are categorized into four classes: \textit{Atris}, \textit{Satris}, \textit{Jigsaw}, and \textit{Random}, based on the shapes of the items.
The \textit{Jigsaw} instances contain convex polygons with edges in random directions as items and convex containers.
The \textit{Random} instances contain arbitrary polygons as items and convex containers.
The \textit{Atris} instances contain slightly perturbed polyominoes as items and rectangular containers.
The \textit{Satris} instances contain polygons obtained by shearing polyominoes as items and rectangular containers.
The item values also vary. Some instances have all items with unit value, other instances have values that are small random integers, and other instances have values that are roughly proportional to the area. Solutions to an instance of each class are illustrated in Figure~\ref{f:sols}, and additional details about the challenge can be found in the organizers' survey paper~\cite{survey}.

Our general strategy consists of finding good initial solutions (using integer programming or a greedy heuristic) and subsequently optimizing them with local search. Our strategy shares a geometric greedy approach with the second-place team~\cite{place2}, but they did not use integer programming to obtain initial solutions, and their optimization phase is also different from ours. The third-place team~\cite{place3} uses a hierarchical grid approach. The fourth-place team~\cite{place4} employed a completely different integer programming model and a genetic algorithm.

We describe our algorithms in Section~\ref{s:algorithms} and experimentally analyze their performance using different parameters in Section~\ref{s:parameters}. Section~\ref{s:enginering} presents engineering details focusing on performance. Our solvers were coded in Python and C++ and executed on several desktop and laptop computers, as well as the clusters available in our labs (LIS and LIMOS).

\section{Algorithms} \label{s:algorithms}

We use two different algorithms to compute initial solutions, a preprocessing phase that can be executed beforehand, and a local search phase to improve the solutions.

\subsection{Integer Programming Approach}

A simple idea to solve the challenge problem is to produce a set $V$ of random translations of each item inside the container and then reduce the problem to a type of maximum weight independent set problem in a graph $G=(V,E)$. The vertex set $V$ is a set of translated items. The weight associated to each vertex $v$ is the value of the corresponding item. The edge set $E$ contains two \emph{types of edges}. Two vertices are connected by an edge of the first type if their translated items overlap. Hence, an independent set will not contain overlapping items. The second type of edge enforces that the solution does not have more than the prescribed quantity $q_i$ of a given item $i$. To take this constraint into account, we introduce a clique $C_i$ connecting all the vertices corresponding to item $i$.
If the quantity $q_i$ is $1$, then we need to choose at most one vertex per clique. If all quantities are $1$, then the packing problem (restricted to a finite set of translations) is equivalent to the traditional maximum weight independent set problem.  However, if items have non-unit quantities, then each clique $C_i$ is associated with the quantity $q_i$ of item $i$ and at most $q_i$ vertices of the clique are allowed in the solution.

The combinatorial problem described in the previous paragraph is easily modeled as integer programming with one binary variable per vertex. A type-1 edge $uv$ is modeled as $x_u + x_v \leq 1$ where $x_u$ and $x_v$ are, respectively, the variables associated to the vertices $u$ and $v$. Each clique $C_i$ is modeled as $\sum_{v \in C_i} x_v \leq q_i$. The CPLEX solver~\cite{cplex} can optimally solve graphs with a few thousand vertices obtained from the challenge instances, which is not enough to obtain good solutions using uniformly random translations.

To obtain better solutions, we start from a solution $S$ obtained with the aforementioned method and build a graph $G=(V,E)$ using the same edge rules as before, but for a different set of vertices $V$ constructed as follows. Let $\sigma > 0$ be a parameter and $N$ be a set of the zero vector and random vectors where each random vector has $x$ and $y$ coordinates as Gaussian random variables of mean $0$ and standard deviation $\sigma$. For each translated item in the solution $S$ and for each translation vector in $N$, we create a vertex in $V$ by translating the item accordingly. Translations that are not completely inside the container are removed. We also create vertices in $V$ using uniformly random translations for all items. Edges and cliques are created as before, and the new combinatorial problem is solved with CPLEX. We repeat this procedure multiple times using the previous solution $S$ and reducing the value of $\sigma$ at each step.

This method works well for instances with up to $200$ items. To handle larger instances, we partition the container using a square grid and partition the items equally among the cells. The partition is such that items are grouped by the slope of the longest edge, breaking ties by the slope of the diameter. Each cell is then solved independently. The intuition to group items of similar slope together is that they can often be placed in a way that minimizes the wasted space. Since the values of the items do not seem to be related to the slopes, this approach works well for the challenge instances.

\subsection{Greedy Heuristic}

The greedy heuristic begins by generating an initial list $L$ of $n$ grid points within the container, where typically $1000<n<5000$. List $L$ is then shuffled, and its centroid $c$ is computed and rounded to integer coordinates. The point $c$ is subsequently inserted at the beginning of $L$.

Input items are arranged in a list $I$, sorted by decreasing utility. The utility function, detailed in Section~\ref{s:parameters}, is designed to prioritize items with small area and high value.
Some example of utility functions we use are the item's value, the item's value divided by its area, or the item's value raised to the power of $1.5$ divided by its area.
By starting with high-utility items, the algorithm aims to maximize value while minimizing occupied area. Although this strategy seems intuitive, alternative approaches were explored, such as placing small items at the end of the list to fill remaining spaces as the container becomes full or during local search. However, determining the threshold size for this approach proved challenging, and the idea was ultimately abandoned due to the challenge time constraints.

Regardless of the sorting criteria used, the packing is then constructed by considering items from list $I$ one by one in order.

At each step, we have a current packing and a new item $i$ to pack. We first try to pack $i$ at the first position $g\in L$, namely in the center of the container. If it overlaps a previously packed item, we try some random positions around $g$. If we do not find a valid position for item $i$ around $g$, we move to the next position $g$ in list $L$ and try some random positions around it. Then the routine that we repeat at each position $g$ of $L$ is the following: first try to pack item $i$ at grid position $g\in L$. If there are overlaps with the previously packed items or the boundary of the container, try some random positions around $g$. If no valid position is found for item $i$ around $g$, skip to the next position.

If all positions in $L$ have been considered  without success for item $i$, we skip to the next item in $I$.

When an available position is found for an item $i$, the item placed at this position is fully contained in the container and does not overlap any previously packed item. However, positioning item $i$ at this first available position is not necessarily optimal. We attempt to move item $i$ so that it does not remain in the middle of the empty space in the container.

We select a direction $u$ along which item $i$ is \emph{pushed} until it can no longer advance due to obstructions from either other packed items or the boundary of the container. Pushing an item entails translating it in directions $v$ such that the dot product $v.u$ with $u$ is strictly positive. Since the dot product of the item’s position with $u$ always increases during the process, the possibility of loops is avoided. The push routine terminates when no further movement is possible in any direction $v$.

\begin{figure}[htb]
 \centering
\includegraphics[height=5.5cm]{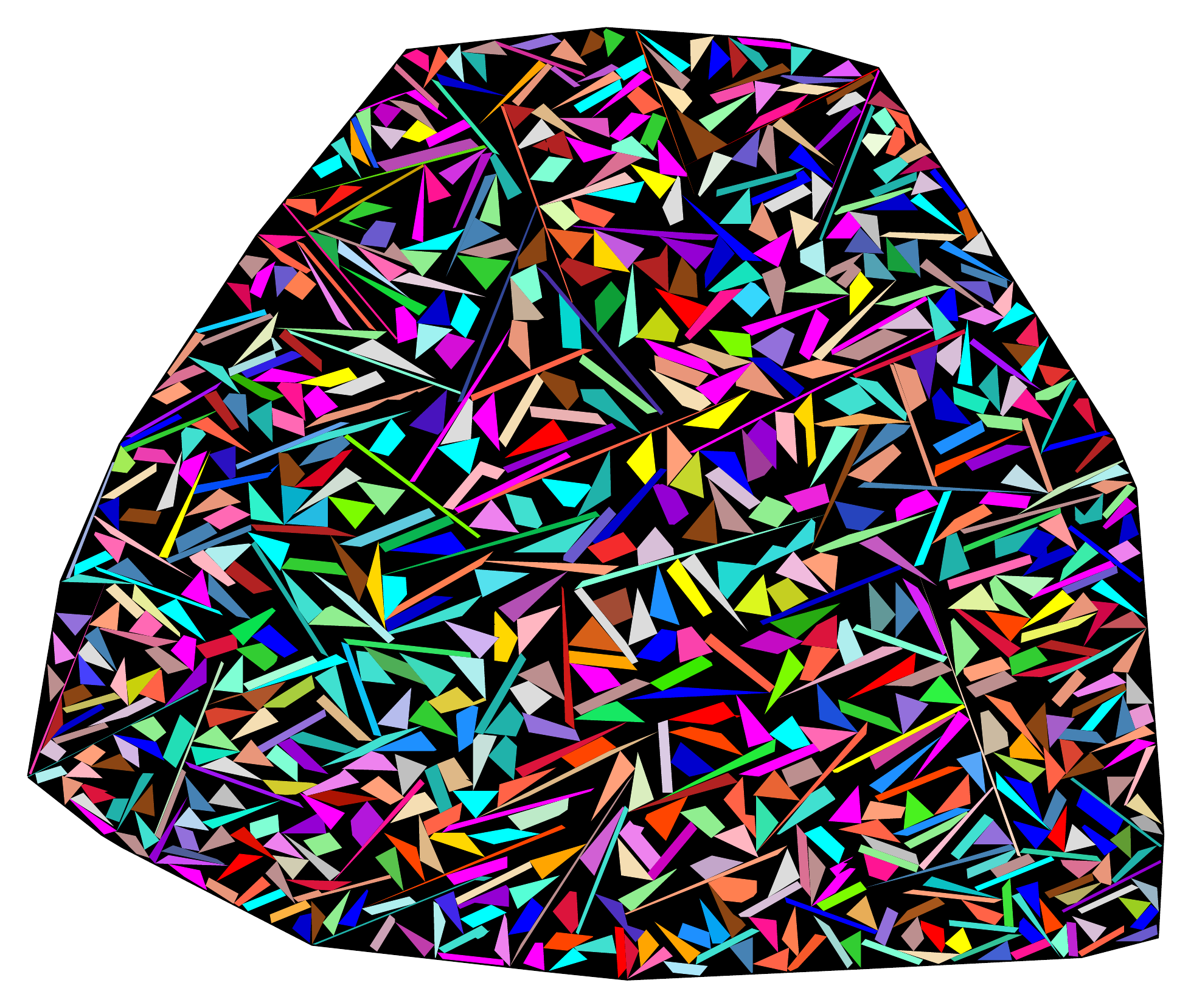}
\hspace{.5cm}
\includegraphics[height=5.5cm]{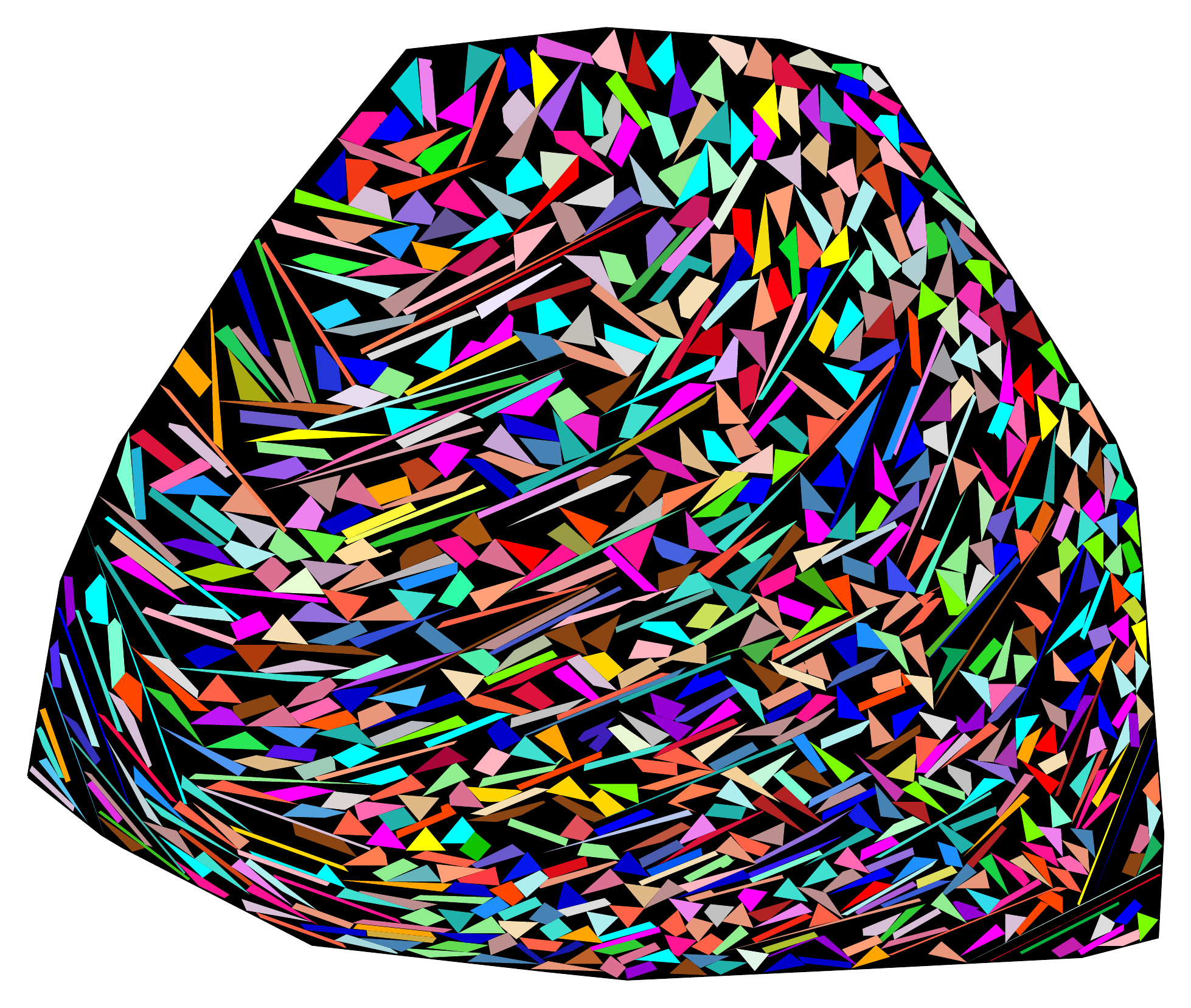}\\
~\hfill(a) Value ratio: $0.82$\hspace{2cm}\hfill(b) Value ratio: $0.85$\hfill~\\
\includegraphics[height=5.5cm]{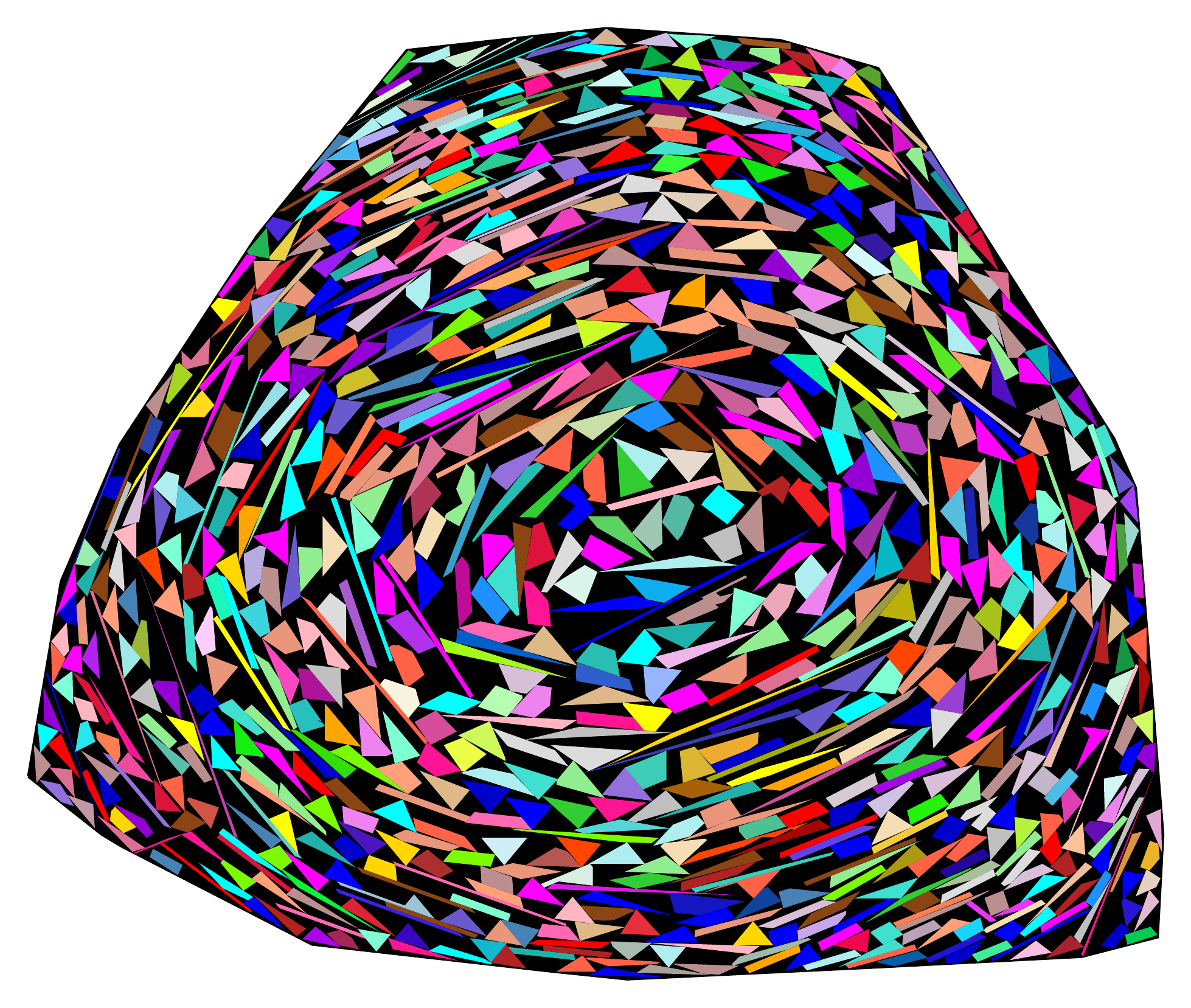}
\hspace{.5cm}
\includegraphics[height=5.5cm]{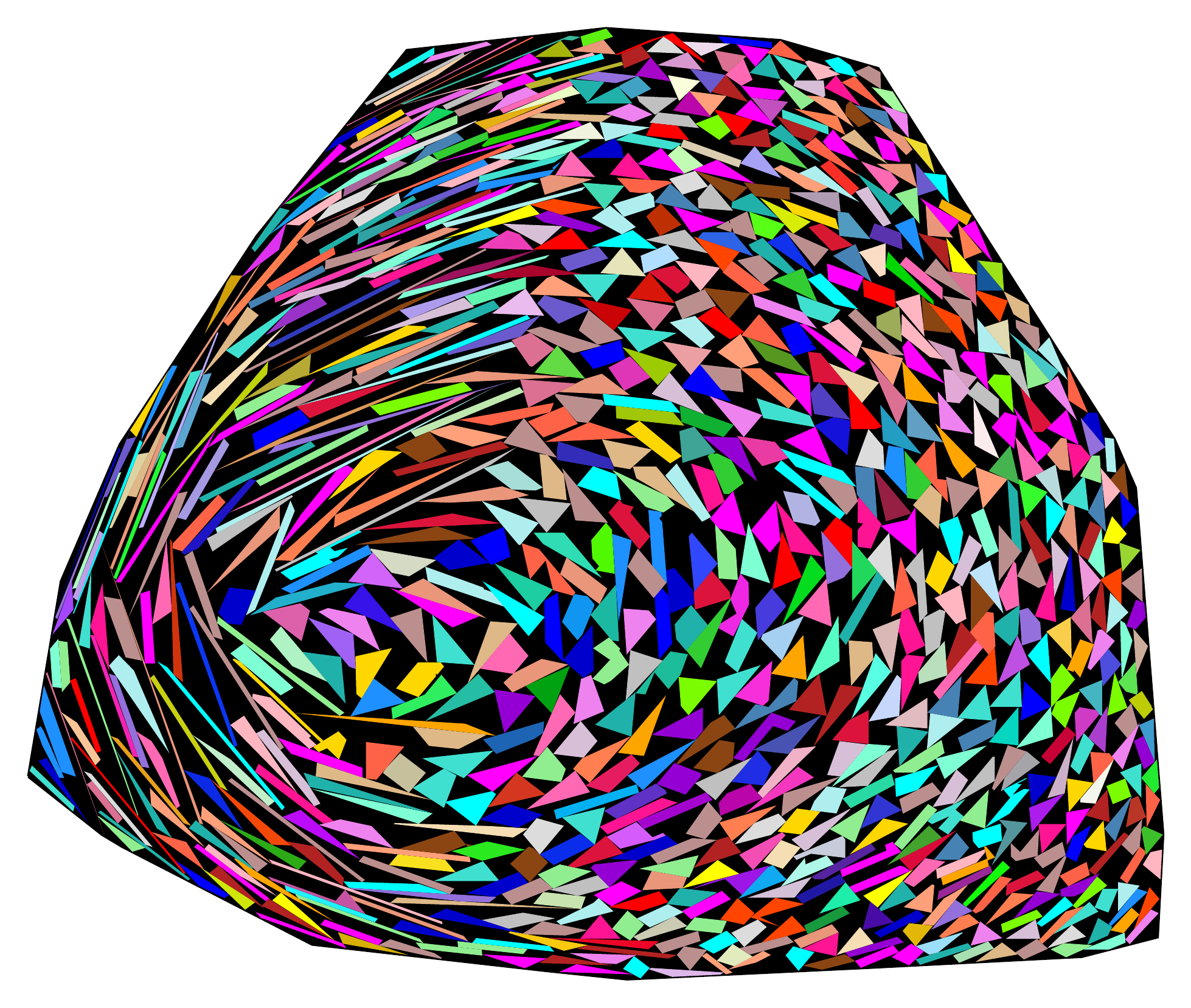}\\
~\hfill(c) Value ratio: $0.89$ \hspace{2cm}\hfill(d) Value ratio: $0.92$ \hfill~\\
  \caption{
  Four solutions for the instance \texttt{jigsaw\_rcf4\_6de1b3b7\_1363} obtained using the greedy algorithm with different strategies to select the direction $u$ to push the items and their value ratios compared to the best solution found during the challenge.
    (a) The direction $u$ is randomly chosen and fixed for all items.
    (b) $u$ is set to be normal to each item’s diameter with negative $y$ coordinate.
    (c) $u$ is again normal to the item’s diameter, but with a random choice of left or right.
    (d) $u$ is normal to the diameter and directed left for skinny items and right for fat items.
  }
 \label{f:greedypackings}
\end{figure}

There are several strategies for selecting the push direction $u$ (see Section~\ref{s:parameters} for details). The most efficient approach is to push the items in a direction normal to their diameter. Alternatively, a separation criterion can be introduced, where skinny items are pushed to the left and fat items to the right.
The greedy algorithm ends when all items in $I$ have either been packed or tested at all positions in list $L$.
Four distinct strategies are illustrated in Figure~\ref{f:greedypackings}.

\subsection{Clusters Preprocessing}

Our greedy heuristic is a first approach which mimics the human approach to solve packing instances but there is a second one. When faced with a packing puzzle, a human would likely first group items into compact clusters before attempting to pack them into the container. In other words, a human would preprocess the items to identify useful clusters.

\begin{figure}[t]
 \centering
 \begin{minipage}[b]{3.7cm}
 \includegraphics[width=3.7cm]{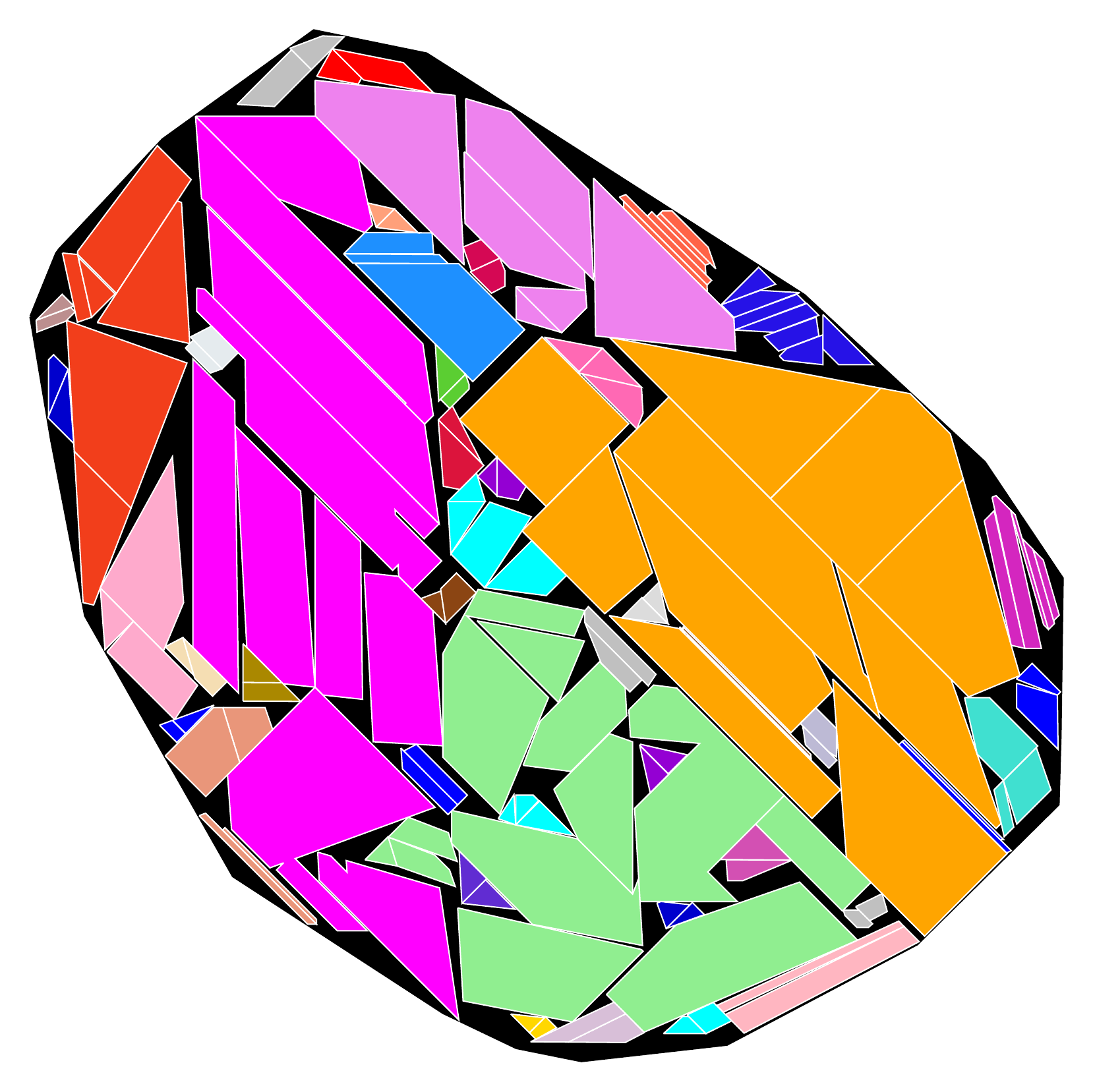}
 \includegraphics[width=3.7cm]{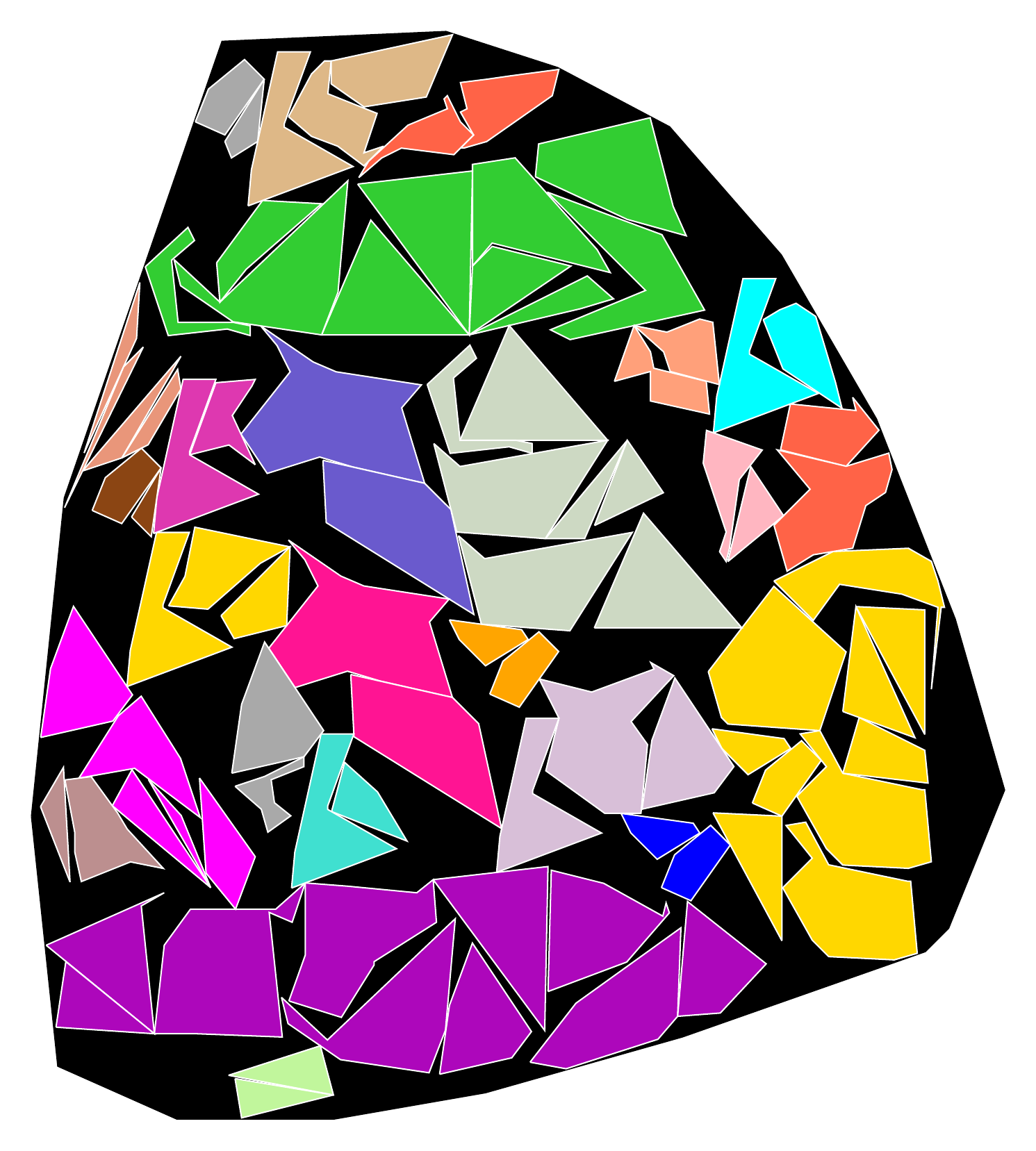}
 \end{minipage}
 \includegraphics[height=8.2cm]{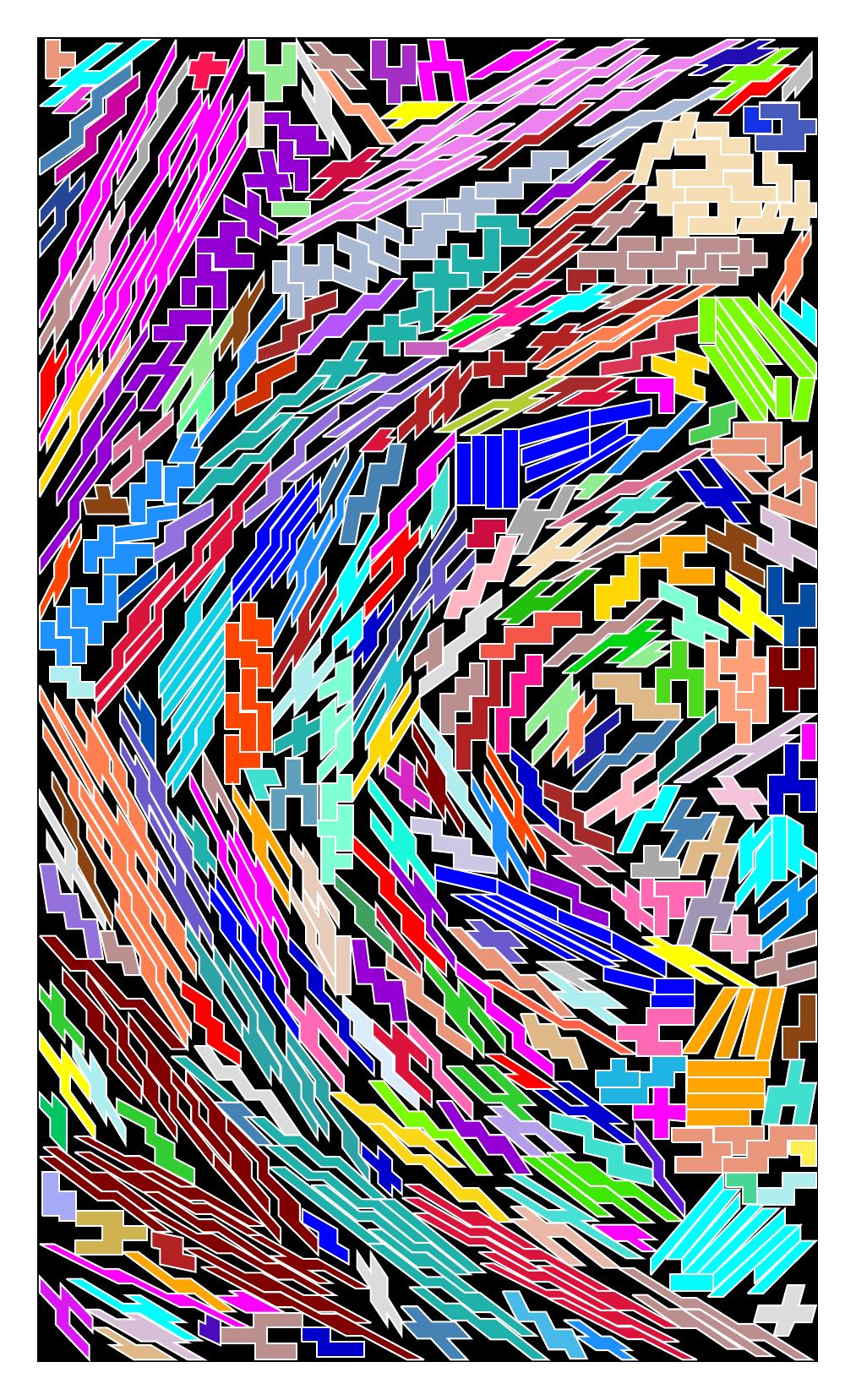}
 \includegraphics[height=8.2cm]{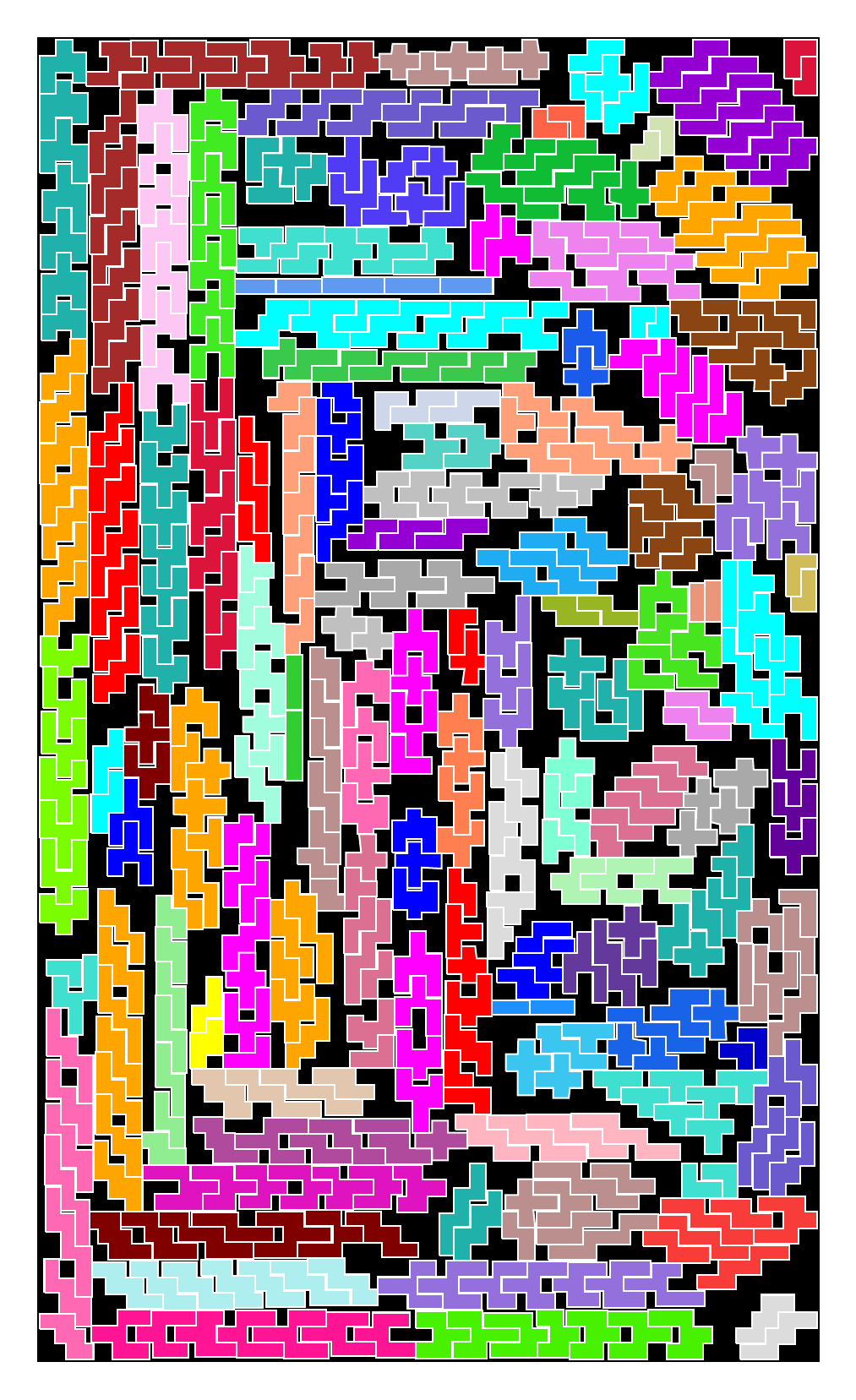}
 \caption{clusters for the instances \texttt{jigsaw\_cf2\_xf42cb20\_670}, \texttt{random\_cf3\_x21f5def\_200},  \texttt{satris1685},  and \texttt{atris1660}. The items of a cluster are drawn with the same color.}
 \label{f:clusters}
\end{figure}

The objective of the preprocessing step is to combine small sets of items into larger groups. The term \textit{cluster} refers to a set of items with fixed relative positions. A cluster functions like an individual item but is a polygonal set rather than a simple polygon (Fig.~\ref{f:clusters}).

\subparagraph{Problem statement.}
Computing clusters is an ill-posed problem until we establish an objective function to define a 'good' cluster. Utility functions from greedy algorithms could serve as candidates, but clusters introduce empty space between items, reducing their utility compared to individual items. For instance, if the utility function is the ratio between the sum of values and the convex hull area, the cluster's utility is often smaller than the utility of its constituent items. Consequently, clusters may seem less competitive and be deprioritized in packing orders. Thus, the primary aim is to design a \textit{cluster utility} function to effectively evaluate cluster efficiency.

Given a known efficient packing of the container, the ratio of the total item value to the container area offers insight into the minimum value density required for improvement. This ratio serves as a threshold to reject low-density clusters (and items). To encourage large cluster usage, we make the cluster utility superlinear in value:
$$ \text{cluster utility}(cluster) = \frac{gauss \cdot value^\alpha \cdot penalty }{area}$$
where
\begin{itemize}
\item $gauss$ is a random Gaussian variable with mean $1$ to diversify clusters upon repeated computation,
    \item $value$ is the sum of item values,
    \item $\alpha$ is an exponent between $1$ and $2$,

    \item $area$ is either the convex hull area or a weighted sum with item areas,
    \item $penalty$ is fixed equal to $1$ except for some cases presenting clusters that we would like to discourage. Let us consider for instance that we have $10$ copies of an axis-parallel rectangle of width $10$ and height $1$. There are two ways to assemble the copies of this item: either in a long bar of size $100 \times 1$ or in a square of size $10 \times 10$. The long bar does not seem a good choice but without penalty, its cluster utility would be equal to the one of the square. Then we penalize skinny clusters from a factor $penalty<1$ computed by using the thickness of the items and the thickness of the cluster.
\end{itemize}

The preprocessing goal is to compute clusters with high cluster utility.

\subparagraph{General strategy.}
 Some technical details depend on item types (polyominoes, sheared polyominoes, convex, non-convex) but we first outline the common pipeline before delving into specific shape classes.

With thousands of items, evaluating all possible pairs for assembly is infeasible. Thus, the first task is to construct a \textit{compatibility graph} connecting items that can potentially be assembled together. But instead of computing a single compatibility graph, we compute multiple graphs, each based on different criteria.
It provides a better control on the cluster generation.

The second step consists in generating clusters from a compatibility graph. We start by generating clusters with two items until a fixed maximum number of items. The items are  generation $1$ and  generation $k+1$ (clusters with $k+1$ items) is built from generations $1$ and $k$. The clusters of generation $k+1$ items are computed by assembling clusters of generation $k$ with items which are compatible to at least one of their own items.
This implies that the items within a cluster of generation $k$ form a connected component of size $k$ in the compatibility graphs. If the average degree of the compatibility graphs is larger than a small constant, it becomes impractical to attempt all possibilities. A crucial requirement is to manage the exponential growth of the number of potential clusters.
We achieve this goal in all generations by keeping only a fixed number $m$ of clusters per item. The selection is done according to the cluster utility function. For each item, we keep the $m$ clusters of highest cluster utility.

\subparagraph{Building the compatibility graphs.} If the number of items is small, then we use a unique compatibility graph which is a clique on all items. If the number of items is large, a quadratic number of compatibilities is too large. Instead we compute several  compatibility graphs of smaller size:
\begin{itemize}
    \item The first compatibility graph is denoted \textit{Rand}. We first define a partition of the set of items in subsets of fixed size (e.g., $100$). The compatibility graph $Rand$ is the union of the  cliques of this partition subsets.
    \item A second compatibility graph is denoted \textit{Skinny}. This graph connects the items whose ratio diameter/width is larger than a constant (e.g., $3$). It connects the items whose longest edges have roughly the same direction.
    \item  A third graph is called \textit{Concav}. It connects some items with concavities with some items whose shape could allow us to  fill the holes.
\end{itemize}

The three previous graphs are considered for items with edges in arbitrary directions. For specific instances like Atris and Satris, other graphs can be computed.
\begin{itemize}
    \item Satris items are sheared polyominoes. We build a compatibility graph called \textit{Shear} by taking  the directions of the edges of the items into account. Each item has two dominant directions. We group the items with close dominant directions and take as  \textit{Shear} the union of the cliques of these groups.
    \item Atris items are slightly perturbed polyominoes but the perturbations are sufficiently small to allow us to encode each shape by its contour path i.e a word on the alphabet $\{ (1,0),(0,1),(-1,0),(0,-1)\}$. Then the complementarity of  item pairs can easily be checked by word combinatorics. Nevertheless, the instances of Atris use only a small number of shapes (bars, crosses, L, shapes in Y and waves). 
    The classification of the items allows us to build a compatibility graph that links all elements of a given class to all the elements of a class with a complementary shape. 

\end{itemize}

\subparagraph{Assembly routine.}
We employ two distinct procedures to assemble an item into a cluster:
\begin{itemize}
\item Grid-based approach:
        Center the cluster on a 2D grid containing between $100$ and $1000$ points.
        Translate the item to each grid position.
        For each position, check if the item overlaps the cluster. If not, compute the cluster utility of the union of the cluster and the item at that position.
\item Vertex-based approach:
        Consider all pairs of an item vertex and a cluster vertex.
        Adjust the relative positions of the cluster and the item so that the item vertex is translated onto the cluster vertex.
        If there are no overlaps, then compute the new cluster utility.
\end{itemize}
We may use either or both methods. Ultimately, we retain the best cluster considered throughout the process.
The overall process is the following. First, we build the chosen compatibility graphs. For each graph, we generate clusters of $k$ items by using the assembly routine and keeping only the best clusters containing each item.

When the cluster preprocessing step is enabled in our greedy algorithm, we replace the utility function used for sorting items with the cluster utility function, which is then used to sort all available clusters. Note that when a cluster is packed, the number of copies of its items is reduced, potentially making other clusters unavailable.

Unsurprisingly, clusters work particularly well for the \texttt{atris} instances, where items resemble polyominoes.
During the CG 2024 competition, our cluster-based approach faced scalability limitations: computation time was not adequately controlled, rendering the strategy insufficiently robust for large instances. As a result, clusters were only sparingly employed in our overall solution, highlighting the need for further work on the subject.

\subsection{Local Search}

To improve the quality of solutions produced by the previous approaches, we employ a local search optimization scheme, which iterates the following routine until a specified time limit is reached.
At each iteration, one of two routines is randomly selected: a \textit{fill} routine or a \textit{dig} routine. The fill routine, akin to a greedy heuristic, attempts to pack all unpacked items into the grid by testing various grid positions and random placements.

The strategy of the dig routine is straightforward: it selects a point $v$ within the container and attempts to push the packed items away from this point to free up space.

First, the routine randomly chooses the point $v$, either in the interior or at a vertex of the container. Next, the packed items are sorted by the distance of their centroids from $v$, in descending order. Each packed item $i$, with centroid $c_i$, is then pushed in the direction of $c_i-v$, aiming to clear the space around $v$.

Once the items have been repositioned, the routine attempts to pack the remaining unpacked items around $v$ by using nearby grid points $v$ from the set $L$ and several random integer points around each grid point. In cases where packing a new item requires removing others due to overlap, the routine computes the benefit of removing the conflicting items by comparing their values. If the benefit of replacing them with the new item is non-negative, the replacement is made.

To speed up the procedure for larger instances, a parameter is introduced to restrict the pushing operation to items near $v$, either by defining a radius or by setting a maximum number of items to push.

\section{Parameter Analysis} \label{s:parameters}

\begin{figure}[p]
 \centering
\includegraphics[height=3.5cm]{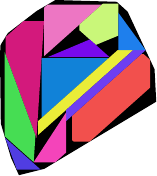}\hfill
\includegraphics[height=3.5cm]{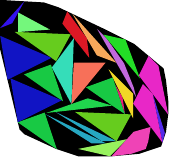}\hfill
\includegraphics[height=3.5cm]{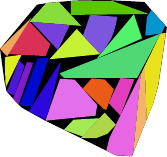}\hfill
\includegraphics[height=3.5cm]{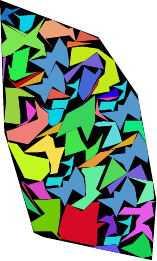} \vspace{2mm}
\includegraphics[height=3.3cm]{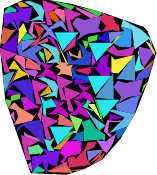}\hfill
\includegraphics[height=3.6cm]{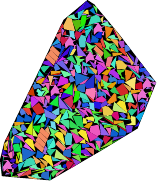}\hfill
\includegraphics[height=3.3cm]{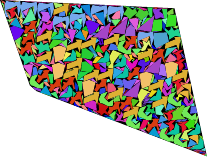}\hfill
\includegraphics[height=3.6cm]{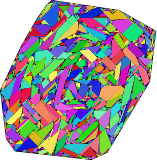} \vspace{2mm}
\includegraphics[height=5.3cm]{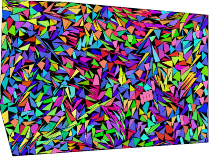}\hfill
\includegraphics[height=5.3cm]{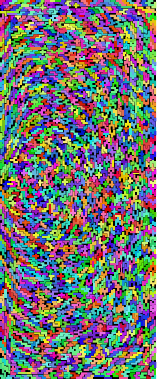}\hfill
\includegraphics[height=5.3cm]{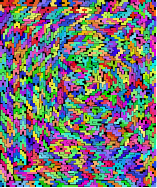} \vspace{2mm}
\includegraphics[height=5.3cm]{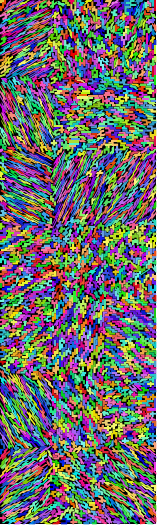}\hfill
\includegraphics[height=4.7cm]{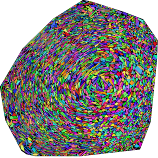}\hfill
\includegraphics[height=5.3cm]{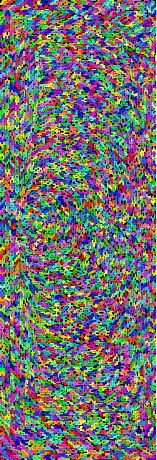}\hfill
\includegraphics[height=4.7cm]{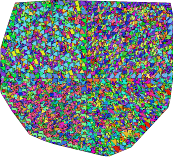}\hfill
  \caption{Our best solutions to the first 15 rows of Table~\ref{t:instances}, in order.}
 \label{f:bestpackings}
\end{figure}

In order to compare the different parameters, we consider $18$ instances of various sizes, as shown in Table~\ref{t:instances}. Throughout, we refer to the ratio between the value of the solution and the value of the best solution in the challenge as the \emph{value ratio} (notice that the competition score is the square of the value ratio). The best solutions we found to the first $15$ table instances are shown in Figure~\ref{f:bestpackings}.

\begin{table}[t]
\begin{center}
\begin{tabular}{l|cccc|c}
Instance & IP & IP + LS & Gr & Gr $+$ LS & Our best\\
\hline
\texttt{jigsaw\_cf4\_273db689\_28} & $\mathbf{1}$ & $\mathbf{1}$ & $.900$& $.950$& $1$ \\
\texttt{random\_cf1\_64ac4991\_50} & $\mathbf{.926}$ & $\mathbf{.926}$ & $.851$ & $.925$ & $.926$\\
\texttt{jigsaw\_rcf4\_7702a097\_70} & $.959$ & $\mathbf{.982}$ & $.867$  & $.924$ & $1$\\
\texttt{random\_cf4\_50e0d4d9\_100 } & $.943$ & $\mathbf{.972}$ & $.877$  & $.941$ & $1$\\
\texttt{random\_cf3\_x21f5def\_200} & $.950$ & $\mathbf{.967}$ & $.893$ & $\mathbf{.967}$ & $1$\\
\texttt{random\_rcf1\_340f4443\_500 } & $.952$ & $.960$ & $.932$ & $\mathbf{.981}$ & $.984$\\
\texttt{random\_rcf3\_x7651267\_1000} & $.927$ & $.970$ & $.926$ & $\mathbf{.980}$ & $1$\\
\texttt{jigsaw\_rcf4\_6de1b3b7\_1363} & $.934$ & $.948$ & $.928$ & $\mathbf{.980}$ & $1$\\
\texttt{random\_cf1\_x51ab828\_2000} & $.937$ & $.945$ & $.892$ & $\mathbf{.950}$ & $.961$\\
\texttt{atris2986} & $.910$ & $.931$ & $.881$ & $\mathbf{.948}$ & $.988$\\
\texttt{atris3323} & $.893$ & $.918$ & $.866$ & $\mathbf{.951}$ & $1$\\
\texttt{satris4681} & $.919$ & $.930$ & $.886$ & $\mathbf{.937}$ & $1$\\
\texttt{jigsaw\_cf1\_4fd4c46e\_6548} & $.944$ & $.944$ & $.955$  & $\mathbf{.963}$ & $.975$\\
\texttt{atris7260} & $.886$ & $.902$ & $.872$& $\mathbf{.920}$ & $.974$\\
\texttt{random\_cf3\_x4b49fe2\_10000} & $.924$ & $.938$  & $.889$ & $\mathbf{.963}$ & $1$\\
\texttt{satris15666} & $.923$ & $\mathbf{.951}$ & $.877$ & $.910$ & $1$\\
\texttt{jigsaw\_cf1\_203072aa\_32622} & $.732$ & $.876$ & $.972$ & $\mathbf{.973}$ & $.985$\\
\texttt{atris41643} & $.513$ & $.803$ & $.854$ & $\mathbf{.877}$ & $.911$
\end{tabular}
\end{center}
\caption{Value ratio of several instances using integer programming (IP) or greedy (Gr), both before and after 24 hours of local search (LS). The last column shows the best value ratio that the Shadoks team obtained in the competition.}
\label{t:instances}
\end{table}

\begin{figure}[p]
 \centering
 \hspace{-0.2cm}
 \includegraphics[width=11.9cm]{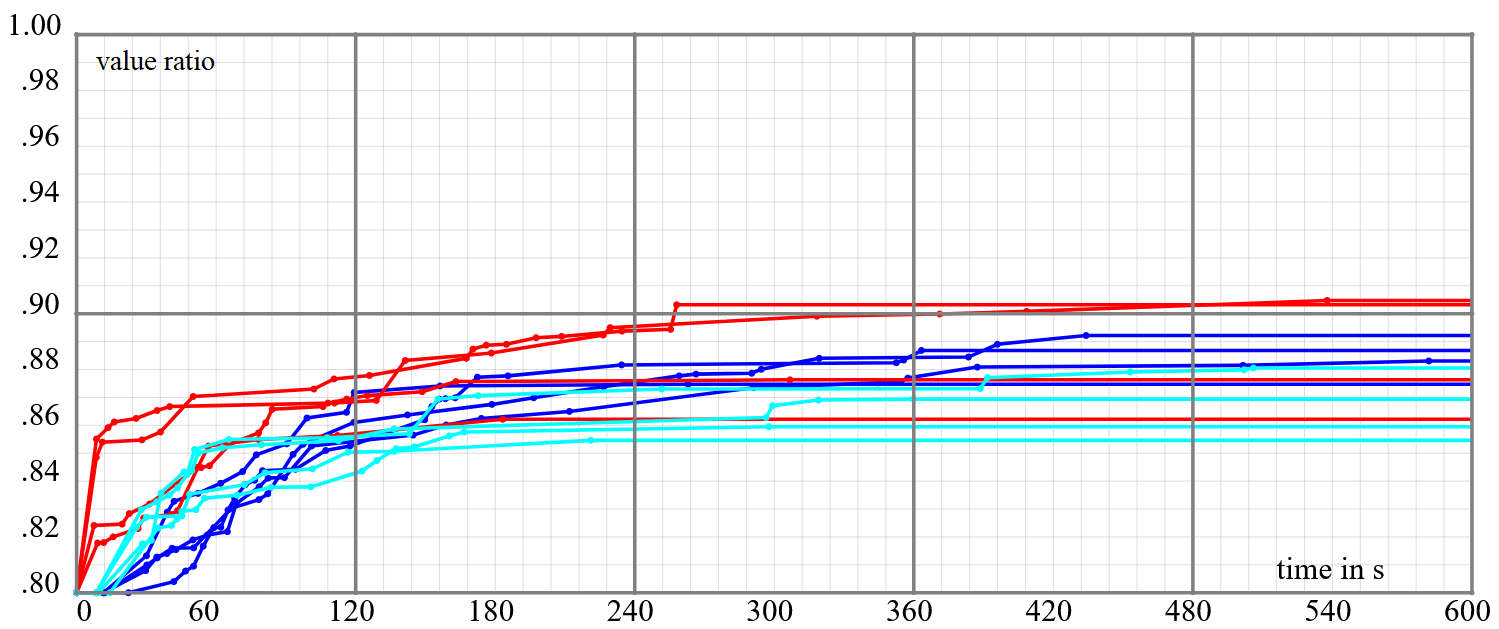}
  \texttt{random\_cf4\_50e0d4d9\_100 }
  \includegraphics[width=12cm]{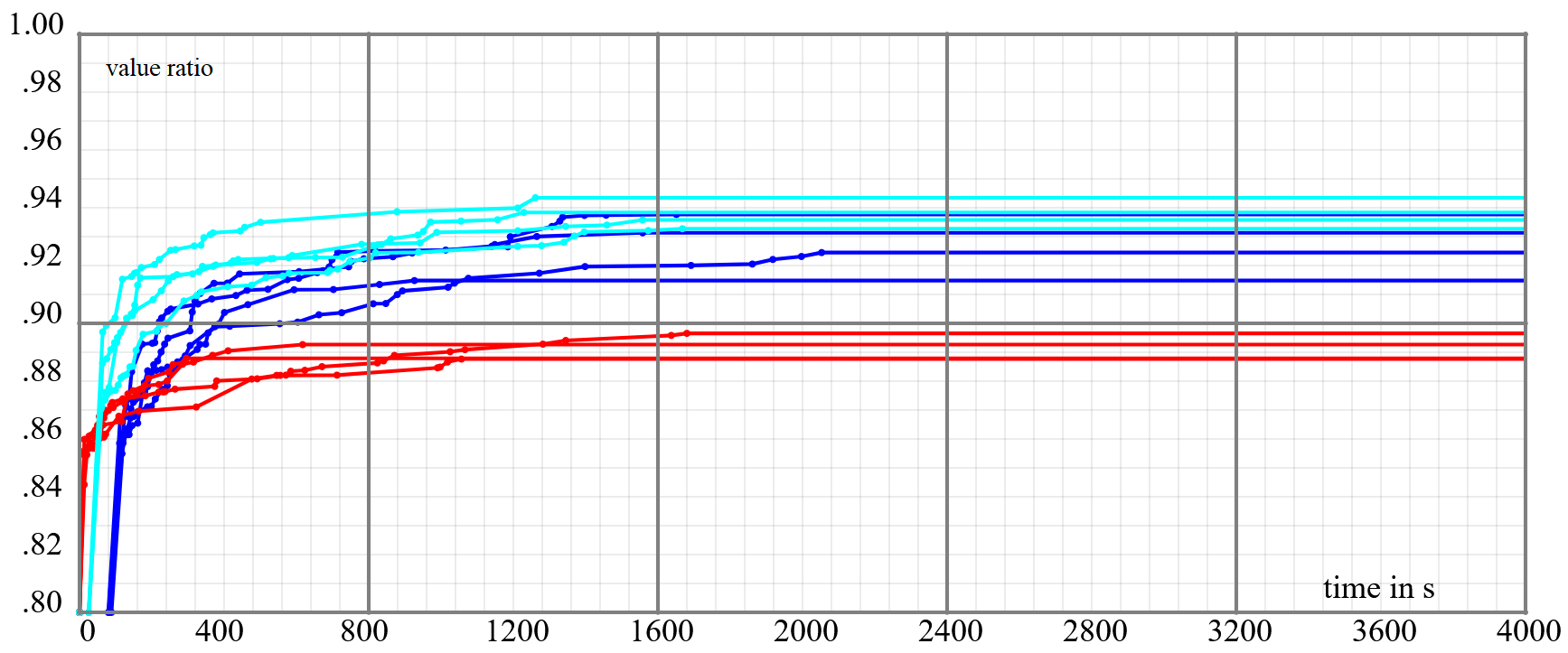}
  \texttt{random\_cf3\_x21f5def\_200}
  \hspace{0.1cm}
 \includegraphics[width=12.2cm]{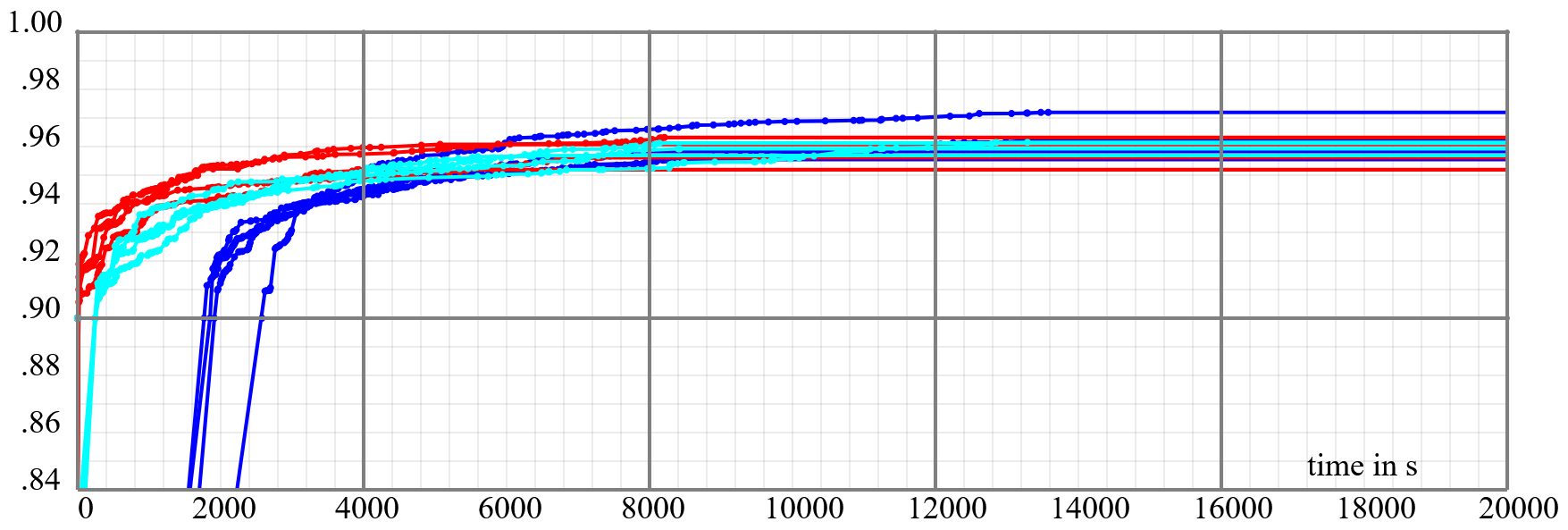}
 \texttt{random\_rcf3\_x7651267\_1000}
 \caption{The value ratio over time for the instances \texttt{random\_cf4\_50e0d4d9\_100},  \texttt{random\_cf3\_x21f5def\_200}, and \texttt{random\_rcf3\_x7651267\_1000}
 for $12$ independent executions of the greedy algorithm and a subsequent local search. Red curves are computed without cluster preprocessing, while the light blue and dark blue curves respectively include a preprocessing of clusters with at most $5$ and $10$ items.}
 \label{f:time}
\end{figure}

Figure~\ref{f:time} illustrates the evolution of the value ratio when applying the cluster preprocessing step, followed by the greedy algorithm and local search until stabilization, using standard parameters for three instances with $100$, $200$, and $1000$ items, respectively. The results indicate that while clustering can sometimes improve performance, it may also have a negative impact in certain cases.

\begin{figure}[p]
 \centering
\includegraphics[width=13cm]{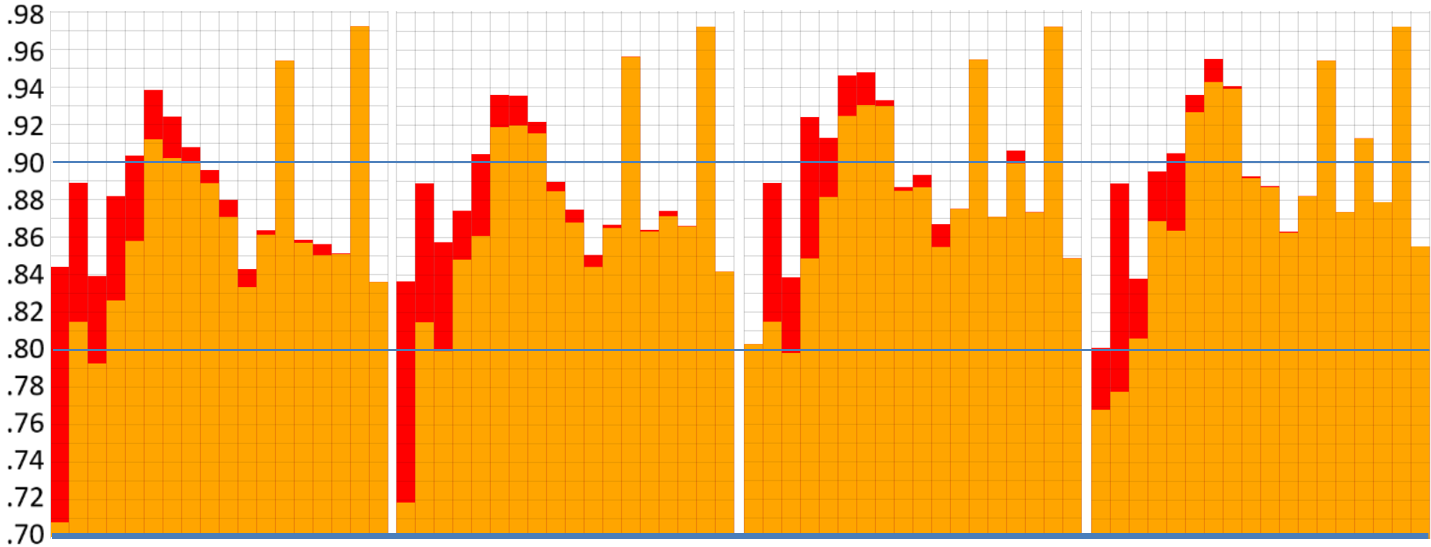}
\centerline{\hspace{0.5cm}500 positions \hspace{0.75cm} ~1000 positions \hspace{.75cm} 2500 positions \hspace{.75cm} 5000 positions}
 \caption{Value ratio of the greedy algorithm (orange) for $500$, $1000$, $2500$ and $5000$ grid positions, for the 18 instances in Table~\ref{t:instances}. In red, we show the value attained after 1000 seconds of local search.}
 \label{f:positions}
\end{figure}

\begin{figure}[p]
 \centering
\includegraphics[width=13cm]{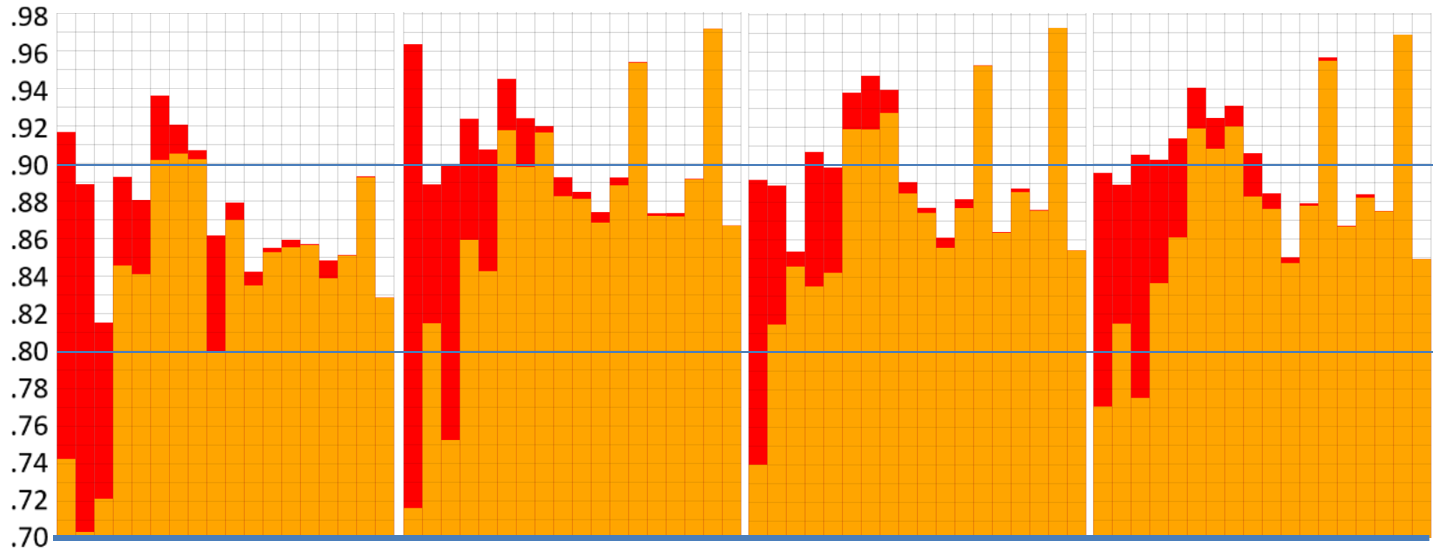}

\centerline{\hspace{1.1cm} $value$  \hspace{1.7cm} $value/area$ \hspace{1cm} $value^{1.5}/area$\hspace{0.7cm} $(1+t) value / area$}

 \caption{Value ratio of the greedy algorithm (orange) using different utility functions, for the 18 instances in Table~\ref{t:instances}. $t$ is the ratio between the length of the item diameter and the width of the item in the direction  normal to the diameter. In red, we show the value attained after 1000 seconds of local search.}
 \label{f:utility}
\end{figure}

\begin{figure}[p]
 \centering
\includegraphics[width=13cm]{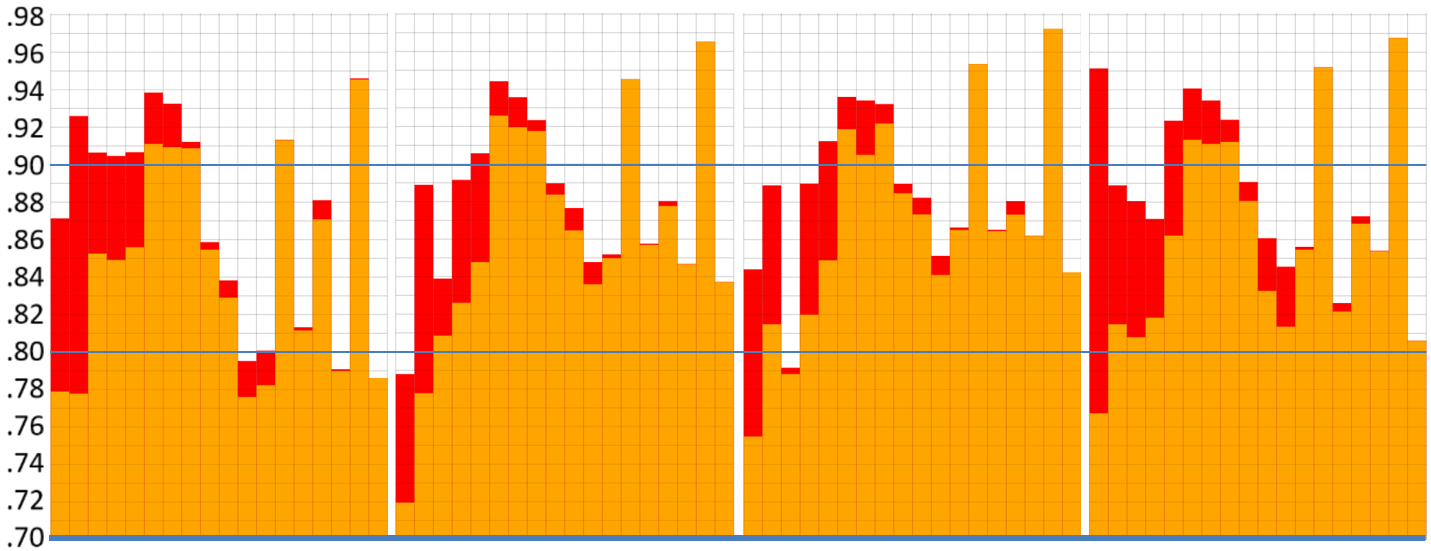}
\centerline{\hspace{1.4cm} random direction \hspace{0.3cm} diameter normal \hspace{0.4cm} diameter + width  \hspace{0.1cm} diameter + longest edge\hspace{0.5cm}}

 \caption{Value ratio of the greedy algorithm (orange) using different push strategies for the 18 instances in Table~\ref{t:instances}. In red we show the value attained after 1 hour of local search.}
 \label{f:choice}
\end{figure}

We now analyze the parameters of the greedy heuristic:
\begin{itemize}
    \item The number of grid positions and random integer positions considered around each point varies. Typically, we use between $500$ and $3000$ grid positions, along with $3$ to $10$ random positions per grid point. Figure~\ref{f:positions} compares different parameter settings and shows that using more than $1000$ grid positions offers no significant improvement.

    \item The utility function used to sort the items $I$ may use various criteria: the item's value, the item's value divided by its area, or the item's value raised to the power of $1.5$ divided by its area, with or without additional weighting factors. For the instances in Table~\ref{t:instances}, Figure~\ref{f:utility} shows the value ratio after the greedy phase and again after optimization, for different utility functions.

    \item The direction $u$ in which an item is pushed can be selected in different ways. It may be chosen randomly (strategy 1) or as a normal to the item's diameter, with a random choice of which side to push (strategy 2). If the item is skinny (with a diameter-to-width ratio greater than $3$), then it is pushed to the left along the normal to its diameter; if the item is fat (ratio less than $3$), then it is pushed to the right (strategy 3). Additionally, items may be pushed in the direction normal to their longest edge if they are fat (strategy 4). Figure~\ref{f:choice} shows the average results over 10 runs for each instance from Table~\ref{t:instances}. According to our experiments, the best strategy is strategy 3. Additional experiments could have been done by using directions other than just left and right as for instance up and down but we assumed that the results would be similar for the instances of the challenge due to the shapes of the containers.
\end{itemize}

Since our experiments to identify the optimal utility function yielded inconclusive results,  a pragmatic strategy is to reduce the grid resolution to accelerate the greedy algorithm’s execution in order to  broader test of alternative utility functions. Nevertheless, selecting the best parameters for a given instance remains a non-trivial problem.

\section{Engineering} \label{s:enginering}

We emphasize several critical points that must be carefully implemented in order to obtain efficient code. In both geometric heuristics and local search, the majority of computational time is spent determining whether two items overlap.

\subsection{Fast Crossing Detection}

The items are preprocessed into data structures to minimize the computational time required to check whether two translated items intersect. Each item stores its axis-parallel bounding box. Given the items positions, if the translated bounding boxes do not overlap, further checks are unnecessary.

The data structure also includes a decomposition of the item boundary into monotone polylines, going from left to right (plus a boolean value telling whether the item is above or below the polyline). We call them \textit{chains}. Each chain is also assigned its own bounding box. Convex items have only two chains. Intersections between chains are computed in linear time. If the chains of two items cross, then the  items overlap. In cases where the chains share common points without crossing (such as a shared vertex) special attention is required. These degenerate cases are particularly significant, as they correspond to configurations where the items are positioned to touch but not overlap, which is often the desired outcome.

\subsection{The Grid}

When packing a large number of items, we employ a grid structure where each cell stores a list of items that intersect with it. To determine whether a new item can be placed at a specific position, we first identify the grid cells $C$ that the item crosses. From these cells, we retrieve the list $I_C$ of items intersecting them. To check if the new item $i$ can be placed at the considered position, we only need to test for intersections between $i$ and the items in $I_C$.

After optimizing the intersection tests, it is essential to also implement the push routine efficiently.

\subsection{Push Routine and Issues}

A key component of the greedy heuristic and optimization process is the operation that involves pushing an item $i$, initially placed at position $p$, in a chosen direction $u$ given by an integer vector. Although this algorithm may seem straightforward to implement, there are several pitfalls to avoid.

The push routine is designed to allow items to slide along the boundaries of other packed items and the container. To achieve this, we consider multiple integer vectors $v$ such that the dot product $u \cdot v$ is strictly positive, then translate the item as far as possible along the ray passing through $p$ in direction $v$. If a new valid position is found, we update the position $p$ and test another direction $v$. After testing a few directions $v$ without successfully moving item $i$, we end the push routine.

We first choose the integer vector $u$, for example a normal vector to the diameter of the item. Let $u'$ be a vector perpendicular to $u$. We try vectors $v$ of the form $v=u+\alpha u'$ for integer $\alpha$ from $-8$ to $8$.

The primary computation involves translating the item in the direction of $v$. Here, we distinguish between \textit{pushing}, which allows sliding in other directions as long as the dot product $u \cdot v$ increases, and \textit{translating}, which refers to movement strictly in direction $v$.

The translation in the direction $v$ starts by computing the smallest factor $2^k$ for integer $k$, such that $p + 2^kv$ lies outside the container.
We then repeat the following procedure.
We decrement $k$ until we find a position $p + 2^kv$ where item $i$ can be placed or $k$ becomes $0$. If such a position is found, we update $p$ to the new position $p + 2^kv$. We then repeat the procedure for the same vector $v$ and smaller values of $k$ until $p + v$ is not a valid position.

A problem that may arise is that, since $v$ is an integer vector, its length may be large, potentially leaving empty space between $p$ and $p+v$.
To address this, we continue the process by investigating new positions along $p+2^kv$ for progressively smaller values of $k$ (e.g., $k=-1$, $k=-2$, and so on).
Since only integer translations are allowed, we may need to round the coordinates of $2^k v$ when $k$ is negative. This introduces two potentially significant issues that need to be addressed.
Both issues arise from the fact that division and rounding can alter the direction of the vector.

The first issue happens in the following case. The vector $\textrm{round}(2^k v)$ is valid but since it is the first time that we test this rounded direction, it might happen that the item can be translated by $2 ^r\textrm{round}(2^k v)$ with a large integer $r$. To precise the issue, let us consider for instance an example with $v=(2,3)$. We assume that the item cannot be translated by $v$. Then we round it to $2 ^{-1}v$ which becomes $(1,1)$. It is possible that in this new direction we can translate the item from $\alpha v$ with $\alpha$ going from $1$ to, for instance, $10^8$. If we repeat translations of vector $(1,1)$ until reaching an obstacle, it will take a very long running time.
To avoid this problem, each time that the direction is altered by a rounding, we proceed as if it was a new direction: we set $v$ to $\textrm{round}(2^k v)$ and we recompute the smallest integer $k$, such that $p + 2^kv$ lies outside the container and start decreasing it as before.

The second issue is that $u \cdot \textrm{round}(2^k v)$ may become negative. If that happens, we stop the calculation and test other different directions $v$.
Despite its apparent complexity, the push routine we implemented, which proceeds by discrete jumps and subsequently calls only the overlapping test, is surprisingly efficient.

\section{Conclusion}

The  knapsack polygonal packing problem is extremely complex. In our experience competing in six annual CG:SHOP challenges, finding  solutions that we believe are optimal has never been as hard. After months of work, we still managed to improve several of the smallest competition instances. Comparing the top three teams, there are only $4$ of $180$ instances for which at least two teams were in a tie with the best solution score.

The algorithms presented in this paper are the main ones that we used during the Challenge. With time to step back, these solutions hold numerous opportunities for improvement. There is significant room for enhancement, both in the clustering preprocessing step and in the initialization of a solution. Using clusters generation to select the items placed in each cell of the integer programming approach for large instances is another promising avenue for improvement. Additionally, our local search could be further refined by incorporating more sophisticated search techniques.

A high-level analysis of our results reveals that, for small instances, the combinatorial framework of the integer programming algorithm outperforms the greedy geometric approach. However, as the instances grow larger, the situation reverses. The combinatorial nature of the solutions demands increasingly high computational power, whereas the geometric approach offers fast heuristics that yield solutions of comparable quality.
Better algorithms could be developed by combining the strengths of both approaches.

\bibliography{references}

\end{document}